# Secure Cloud Communication for Effective Cost Management System through MSBE

## Gaurav Raj[1] and Kamaljit Kaur[2]

[1]Asst. Professor, Lovely Professional University, Phagwara, India
`er.gaurav.raj@gmail.com`
[2]Asst. Professor, Lovely Professional University, Phagwara, India
`kamal.aujla86@gmail.com`

## ABSTRACT

*In Cloud Computing Architecture, Brokers are responsible to provide services to the end users. An Effective Cost Management System (ECMS) which works over Secure Cloud Communication Paradigm (SCCP) helps in finding a communication link with overall minimum cost of links. We propose an improved Broker Cloud Communication Paradigm (BCCP) with integration of security issues. Two algorithms are included, first is Secure Optimized Route Cost Finder (S-ORCF) to find optimum route between broker and cloud on the behalf of cost factor and second is Secure Optimized Route Management (S-ORM) to maintain optimum route. These algorithms proposed with cryptographic integrity of the secure route discovery process in efficient routing approaches between broker and cloud. There is lack in Dynamic Source Routing Approach to verify whether any intermediate node has been deleted, inserted or modified with no valid authentication. We use symmetric cryptographic primitives, which is made possible due to multi-source broadcast encryption scheme. This paper outlines the use of secure route discovery protocol (SRDP) that employs such a security paradigm in cloud computing.*

## 1. INTRODUCTION

Efficient solutions to the problem of routing in BCCP [1] can be challenging under the presence of spiteful nodes that could intentionally break the protocol and/or broadcast ambiguous information. Secure routing protocols usually mandate cryptographic authentication *to decrease the autonomy* of attackers to violate rules. We are focusing over the problem of secure communication links and routing based protocols. Lots of Route Discovery Protocols (RDP) based on the Dynamic Source Routing (DSR) has been proposed over cloud network for efficient communication. [2]

RDP used in Broker cloud communication link needs to follow few defined steps as:

1. A source node (Broker/ Cloud) desiring to find a path to destination (Cloud/ Broker) floods a route request (RREQ) packet.
2. Each intermediate node updates its ID / IP address information in RREQ and forwards it to its neighbours.
3. When RREQ packet arrives at the destination node, it sends back a route response (RREP) packet along with the reverse route which is defined in RREP.

4. Secure DSR protocols will make use of cryptographic authentication to facilitate verification and the reliability of the established route.

In most secure Dynamic Source Routing Protocols, intermediate nodes that promote the RREQ cannot identify spiteful alterations to RREQ packets. In some protocols the destination node can detect inconsistencies and drop such requests. In few protocols, only the source node, at the end of the reverse route, may sense irregularities once the RREP reaches the source node. [2]

Here, in-between nodes have to be provided new methods to detect inconsistencies in RREQ to avoid upcoming modified RREQs, which will be failed after wasting lots of execution time and network bandwidth, caused for network delay and increase the possibilities of unwanted traffic. In addition, Due to some of the issues that render early detection difficult, we propose an efficient way out, employing only symmetric cryptographic primitives.

## 1.1. Broker - Cloud Communication Paradigm

BCCP is very helpful to understand the backend communication between client, broker, cloud exchange and clouds. It provide the basic idea of communication but it needs to revise in terms of many aspects like authenticity, data security, SLA, cloud exchange responsibilities etc.

This process needs to be described broadly in terms of responsibilities of each entity.

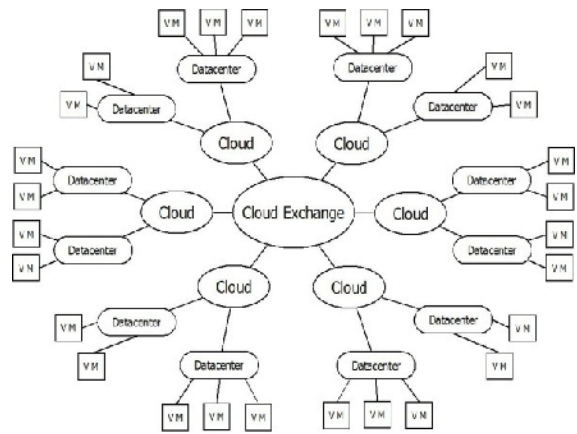

Figure1. Cloud Exchange Scenario

**Client:**

1. Task specifications should be clearly identified
2. Minimum uses and maximum budget also need to be clearly identified.
3. Client should be capable to take decision fast in terms of selecting options provided by brokers.

**Broker:**

1. Need to manage two workflows securely; one is between broker and cloud exchange and other with cloud at the time of task processing.
2. Broker should be capable to transmit the data with cryptographic authentication *to decrease the autonomy* of attackers to violate rules.

3. Broker needs a cost management system to handle commercial services.

**Cloud Exchange:**

1. Need to be updated the information of each connected cloud like its services, cost matrices, route mechanism and few security related information like public keys of each cloud and digital signature etc.
2. It has to be updated with number of free data centres, types of service supported by cloud, SLA terms and conditions regarding usage cost parameters, channel usage cost etc.

**Cloud:**

1. Cloud Coordinator's working and responsibilities need to be defined.
2. Workflow management and load balancing factor need to be handled.
3. Security issues need to be handling with care.

**Data centre:**

1. Overall system efficiency can be improved in a data centre based Cloud with minimal performance overhead using:

   a. power-aware scheduling techniques
   b. variable resource management
   c. live migration
   d. Minimal virtual machine design

## 1.2. Secure DSR Protocols over Secure Cloud Communication Paradigm

In DSR the route discovery process broadcast a route request (RREQ) packet by the source, indicating the source, the destination, a unique sequence number and a hop-limit. In secure DSR protocols the RREQ packets that are forwarded consists of immutable and mutable fields. We shall henceforth modify the immutable fields of an RREQ by adding path cost based on hop count, bandwidth and network delay in *rreq*. The immutable field specifies the source; destination; sequence number; maximum hop counts and also characteristically includes an authentication established by the source. The mutable fields are revised by all intermediate nodes. Characteristically, every intermediate node can introduce some type of authentication to validate the modifications made to the RREQ. [7]

## 2. CLASSIFICATION OF COMMUNICATION LINK

As per the Service with cloud, we classify the link communication in four parts as follows:

1. Service Developer and Service Provider(Developer-Broker Communication Paradigm)
2. Service Client and Service Provider(User-Broker Communication Paradigm)
3. Service Provider and Cloud (Broker-Cloud Communication Paradigm)
4. Cloud and Resource Provider (Cloud-Resource Provider Communication Paradigm)

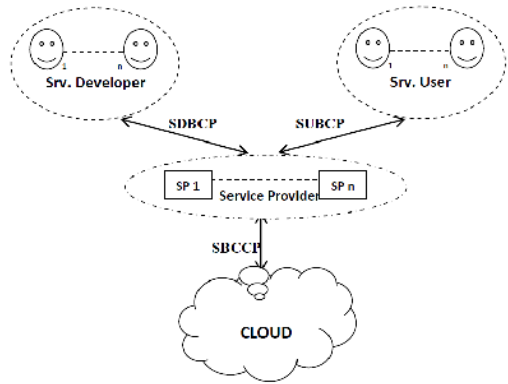

Figure 2. Proposed Secure Cloud Communication Paradigm

## 3. PROPOSED SECURE CLOUD COMMUNICATION PARADIGM [SCCP]

Proposed paradigm can classified as

1. Secure Developer-Broker Communication Paradigm [SDBCP]
2. Secure User-Broker Communication Paradigm [SUBCP]
3. Secure Broker Cloud Communication Paradigm [SBCCP]

### 3.1 Secure Developer-Broker Communication Paradigm [SDBCP]

1. Developer will search for availability of cloud having required resources.
2. Developer will negotiate for resources and SLA terms and conditions.
3. Developer will request for Resources to develop services.
4. Broker will ask to sign SLA and search for Suitable resources.
5. Developer will sign the SLA and request for secure link for communication with broker.
6. Broker will accept the request of secure communication link and will provide the resources to the developer.
7. Broker will handle the control of resource usages by developer and manage the resource allocations, reallocations and withdraw the resources, whenever required to optimise the use of resources.
8. Developer will register the service with broker to gain commercial benefit.

### 3.2 Secure User-Broker Communication Paradigm [SUBCP]

1. Registered service user will search for specific service to process specific task.
2. Broker will request for specifications of task to search the appropriate service for that particular task.
3. User will give the task specifications/ requirements.
4. Broker will provide the information of all available cloud services with economical structure of the service and also suggest best suitable service based on:

    a. Financial analysis
    b. Time analysis.

5. User will select service, will sign the SLA and request for secure communication link with broker.
6. Broker will accept the request and create secure communication link with service user.
7. User will submit task to broker using that secure communication link.
8. Broker will provide information about the usage cost of processing particular task as final result to the user.
9. User will pay according to usage through secure gateways.

### 3.3. Secure Broker Cloud Communication Paradigm [SBCCP]

First we broadly divide the paradigm into three different link communication scenarios

1. Broker – Cloud Exchange Communication[BCEC]
2. Cloud Exchange – Cloud Communication[CECC]
3. Broker – Cloud Communication[BCC]

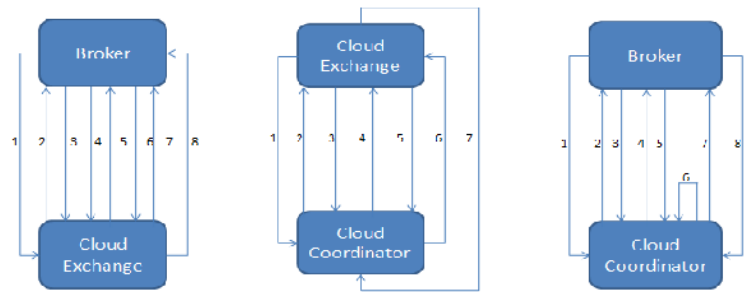

Figure 3. (a) Broker – Cloud Exchange  (b) Cloud Exchange – Cloud Coordinator (c) Broker – Cloud Coordinator

### 3.3.1. Broker – Cloud Exchange Communication [BCEC]

1. Broker will search for available clouds with free data centers in cloud exchange for processing specific task.
2. Cloud exchange will provide the information of available clouds with their statistics and number of available data centers.
3. Broker will take decision on the behalf of requirements and information provided by cloud exchange.
4. Broker will select proper cloud, will sign the SLA and request for secure communication link with cloud exchange.
5. Cloud exchange will accept the request and create secure communication link with broker.
6. Broker will request for authentication key for that required clouds.
7. Cloud exchange will share the authentication key information through secure communication link.
8. Cloud exchange will close the secure link.

### 3.3.2 Cloud Exchange – Cloud Coordinator Communication [CECCC]

Cloud Exchange has the information of all connected clouds, their services, available datacenters, SLA terms and conditions, costing statics etc.

1. Cloud Exchange will request to update above mentioned cloud information at unit time interval.
2. Cloud coordinator will updates all information about services.
3. When Broker will request for authentication key, cloud exchange will inform the cloud about the broker and the task which will require the services of that cloud.
4. Cloud Coordinator will confirm the availability and request to sign SLA.
5. After getting signed on SLA by Broker, cloud exchange will provide it to cloud coordinator.
6. Cloud coordinator provides its authenticated digital signature copy in encrypted form to the cloud exchange.
7. Cloud exchange provides the encrypted digital signature of broker for direct communication in between broker and cloud.

### 3.3.3 Broker – Cloud Coordinator Communication [BCCC]

Broker has encrypted digital signature of cloud which provide the required services.
1. Broker will send the request for service.
2. Cloud coordinator will request for authentication key.
3. Broker will provide encrypted digital signature to the cloud coordinator.
4. Cloud coordinator will decrypt the key and if it will be authenticated then it will request to send the task to process.
5. Broker will send the task encrypted with its digital signature.
6. Cloud coordinator will decrypt the task by using digital signature of broker and send it for processing in datacenter.
7. Cloud coordinator will reply with result in encrypted form and the cost of processing the task.
8. Broker pay for the services using authenticated gateways.

## 4. Efficient Two-Hop Authentication with Secure RDN and Multi – Source Broadcast Encryption (MSBE)

We are implementing two hop authentications in establishing the secure communication link, which requires nodes to only maintain a consistent one-hop topology. This is made feasible by the use of Multi-Source Broadcast Encryption (MSBE) scheme, in combination with maintenance of a Secure Reliable Delivery Neighborhood (S-RDN) by every intermediate node. [2]
To realize S-RDN for every intermediate node, it is to identify nodes within RDN and provide such nodes with a secret Key. If a node $A$ provides a secret key $K_A$ to all nodes in its RDN. All transmissions by $A$ could be encrypted with the secret $K_A$ to make sure that "neighbors" not in the RDN cannot get access to transmissions from $A$.

In the MSBE scheme,
1. A key Distribution Center (KDC) chooses K secret keys $K_1$.... $K_k$, a straightforward one way function $F(\ )$, and a cryptographic hash function $h(\ )$.
2. The public function $F(A) = \{A_1, A_2....... A_m\}$ determines the collection of secret keys allocated to node $A$.
3. Node $A$ is assigned $m\ decryption$ secrets $S_A$, and additionally, $k\ encryption$ secrets $G_A$, where
4.

$$S_A = \{K_{A1}, K_{A2}, \ldots, K_{Am}\} \tag{1}$$

$$G_A = \{K^A{}_j = h(K_j \ // \ A)\}, \qquad 1 \quad j \quad k \tag{2}$$

5. Let U represent the set of all nodes that have been provided with encryption and decryption secret keys, and let $N_A \in U$ be a small subset of "revoked" nodes, which are not provided with the secret keys.

6. Encrypted Broadcast secret key $T_A$ with a subset of its encryption secret keys $G_A' \in G_A$, can be conveyed to all nodes by node $A$ in its RDN *excluding* the "revoked" nodes in the set $N_A$.

7. The particular indices of the encryption keys chosen as part of $G_A'$ for this intention are resolute uniquely, which guarantees that:

      i. Not any of the revoked nodes will have right to use any of the keys in $G_A'$

      ii. All other node in $U \setminus N_A$ will have access to at least one of the secret keys in $G_A'$ with a high possibility.

To transmit the secret key to nodes in the set $U \setminus N_A$ , node $A$ assembles a broadcast message

$$B_A = [N_A \| \{ G_A' (K_A) \} \| M_{TA}], \qquad\qquad (3)$$

$$M_{TA} = h(N_A , G_A' (K_A), T_A ) \qquad\qquad (4)$$

## 4.1. Secure Route Discovery Protocol (SRDP)

1. Source node (Cloud / Broker/ Service Client) will create a Route Request Packet (RREQ) to communicate with destination node, consists of immutable and mutable fields.
   Immutable field have following information-
         $[S_{Addr} \| S_{SeqNo} \| B_{id} \| D_{Addr} \| D_{SeqNo} \| MAX\_Hops ]$
   Mutable field have following information
         [Hop_Count || Path_Cost]

2. Source broadcasts this RREQ to all connected nodes in its S-RDN.

3. If connected Node already received same RREQ again
         Then It drops upcoming RREQ
   Else It updates Hop_Count and Path_Cost fields in mutable field of RREQ.

   Cost is updated by using following equation:
         $Path\_Cost = Path\_Cost + \alpha.1 + \beta.Avl\_Bw + \gamma.Nw\_Delay$     (5)
   where
         Avl_Bw: Available Bandwidth
         Nw_Delay: Network Delay

4. Every Intermediate node updates its routing table and rebroadcast the RREQ in its S-RDN.

5. As the destination node receives the first RREQ, it goes in waiting condition for a unit time.

6. When it collects a number of RREQs, it selects the RREQ with the smallest cost value.

7. Destination sends back a Route Reply Packet (RREP) to the source node via the reverse path defined by selected RREQ.

8. An intermediate node records the previous hops and relays the packet up to source node after receiving RREP.

9. Then source node starts forwarding the task via selected route.

SRDP presumes a suitable KDC System (KDS) for pair wise authentication of intermediate nodes. Every intermediate node maintains a SRDN by providing a group secret key to each node in the RDN. We will represent the one-hop RDN secret keys of a node $A$ offers to all nodes in the set $n_A$(the RDN of $A$) by $K_A$, and the broadcast secret key, which is *confined* from all nodes in the set $n_A$ by $T_A$.

### 4.1.1 RREQ Broadcast

We consider a set-up where a source $S$ requests to find a path to a destination $D$. The source creates a RREQ packet with immutable fields as

$$rreq = [S_{Addr} \| S_{SeqNo} \| B_{id} \| D_{Addr} \| D_{SeqNo} \| MAX\_Hops ]$$

and mutable fields as

$$[Hop\_Count \| Path\_Cost]$$

where $S_{seqNo}$ is a sequence number and MAX_Hops is the maximum hop count. The node $S$ now broadcasts an RREQ packet

$$RREQ_0 = [S_{Addr} \| S_{SeqNo} \| B_{id} \| Hop\_Count \| Path\_Cost \| K_S([rreq \| M_0 \| h_0])]$$
$$h_0 = h(rreq, K_{SD})$$

$$M_0 = h(rreq, h_1, T_S), \text{ where } h_1 = h(h_0) \qquad (6)$$

where $K_{SD}$ is a secret shared between $S$ and $D$. $h_0$ is a Hashed Message Authentication Code (HMAC) designed for authentication by the destination node, $M_0$ is HMAC designed for authentication by two-hop nodes. All fields of the RREQ packet are encrypted by the one-hop secret key of $S$. A node $A$ one hop from $S$ decrypts the RREQ packet and broadcasts

$$RREQ_1 = [A_{Addr} \| A_{SeqNo} // B_{id} \| Hop\_Count \| Path\_Cost // K_A([rreq // (A) // M_0 // M_1 // h_1])]$$

$$M_1 = h(rreq, (A), h_2, T_A), \text{ where } h_2 = h(h_1) \qquad (7)$$

A node $B$ one-hop away from $A$ and two-hops away from $S$ decrypts the RREQ, verifies that $h_1$ sent by $A$ is consistent with the HMAC $M_0$ appended by $S$. Having verified $M_0$, node $B$ strips off $M_0$ and appends an HMAC $M_2$ for verification by nodes two hops downstream of $B$. Thus $B$ broadcasts

$$RREQ_2 = [B_{Addr} \| B_{SeqNo} // B_{id} \| Hop\_Count \| Path\_Cost // K_B([rreq // (A,B) // M_1 // M_2 // h_2])]$$

$$M_2 = h(rreq, (A,B), h_3, T_B), \text{ where } h_3 = h(h_2)$$

**RREP:**

When the node $D$ receives the RREQ packet with a per-hop hash value $h_i$ and $i$ nodes in the path, $D$ can verify that $h_i$ is consistent with $h_0$, which $D$ can evaluate as it is based on a secret $K_{SD}$ shared between $S$ and $D$. We take the assumption that RREP reached the destination via path $(A, B, \ldots , W,X, Y, Z)$. The RREP raised by $D$ *converted in* the form

$$rrep = [S_{Addr} \| S_{SeqNo} \| D_{Addr} \| D_{SeqNo} \| (A,B, \ldots , W,X, Y, Z)].$$

After that the RREP packet has been unicast by node $D$.

$$RREP_0 = [D_{Addr} \| D_{SeqNo} \| K_D([rrep \| q_0 \| M_{DY} ])$$

$$M_{DY} = h(rrep, q_1, K_{DY}), \text{ where } q_1 = h(q_0)$$

It should be Notify that the RREP packet has two Hashed Message Authentication Codes - $q_0$ for verification by the source at the end of the RREP, and $M_{DY}$ for authentication by a node $Y$ two hops away in the RREP path. [2]

The RREP packets relayed by nodes $Z$ and $Y$ take the form

$RREP_1 = [Z_{Addr} \parallel Z_{SeqNo} \parallel K_Z([rrep \parallel q_1 \parallel M_{DY} \parallel M_{ZX}])$
$M_{ZX} = h(rrep, q_2, K_{ZX})$, where $q_2 = h(q_1)$

$RREP_2 = [Y_{Addr} \parallel Y_{SeqNo} \parallel K_Y ([rrep \parallel q_2 \parallel M_{ZX} \parallel M_{YW}])$

$M_{YW} = h(rrep, q_3, K_{YW})$, where $q_3 = h(q_2)$

## 4.2. SRDP Implementation over Secure Cloud Communication Paradigm

This paper improves Effective Cost Management System (ECMS) by using Secure Route Discovery Protocol (SRDP) by implementing Secure Dynamic Source Routing (S-DSR). The ECMS includes two Algorithms:

1. Secure Optimum Route Cost Finder (S-ORCF)
2. Secure Optimum Route Maintenance (S-ORM).

We also improve Multiple Metrics for Path Cost (MMPC), which integrates the cost based on [12]:

1. hop count of intermediate nodes
2. channel bandwidth between pair of communicating nodes
3. Network delay due to congestion or any network failure

These cost metrics define the path cost used for route selection in S-ORCF.[13] A node, that detects a link break due to lack of response from a neighboring node, calls S-ORM which will send a Route Error Packet (REP) to the source node via reverse route define in REP as follows:

$REP = [S_{Addr} \parallel S_{SeqNo} \parallel D_{Addr} \parallel D_{SeqNo} \parallel K_{SD}([ErrorCode]) \parallel (A,B, \ldots , W,X, Y, Z)]$

While receiving the Route Error Packet, the source node initiates a new round of Route Request Packet by calling S-ORCF. [1]

### 4.2.1 Effective Cost Management System (ECMS)

### Algorithm1: Secure Optimum Route Cost Finder (S -ORCF)

Algorithm between Broker and Cloud ($v_i$ and $v_j$) with Hops count, bandwidth and Network delay factors.

**Input:** Three matrices $M_{HC}(v_i, v_j)$ for minimum hops count and $M_{BW}(v_i, v_j)$ for maximum average bandwidth and $M_{ND}(v_i, v_j)$ for minimum average Network delay, where
$(v_i, v_j) \in$ E and Set $(\alpha, \beta, \gamma) = (1, 0.1, 1)$

**Output:** Minimum cost path according to priority factor

1. Select the Cost factor to minimize link cost of communication link from Broker to Cloud.
   i. Hops Count(1 for $\alpha$ ,0 for $\beta$ and 0 for $\gamma$ )
   ii. Bandwidth(0 for $\alpha$ ,0.1 for $\beta$ and 0 for $\gamma$ )
   iii. Network Delay(0 for $\alpha$ ,0 for $\beta$ and 1 for $\gamma$ )
   iv. Hops Count and Bandwidth(1 for $\alpha$ ,0.1 for $\beta$ and 0 for $\gamma$ )
   v. Bandwidth and Network Delay(0 for $\alpha$ ,0.1 for $\beta$ and 1 for $\gamma$ )
   vi. Hop Count and Network Delay(1 for $\alpha$ ,0 for $\beta$ and 1 for $\gamma$ )
   vii. Hops Count, Bandwidth and Network Delay(1 for $\alpha$ ,0.1 for $\beta$ and 1 for $\gamma$ )
2. Case (i) Select the path with SRDP using Minimum Link HopsCount [$M_{HC}(v_i, v_j)$ ].
   If more than one path has same HopsCount
   Then Go to case 7 for selecting path with maximum HopsCount-Bandwidth-Delay Product (HBDP)
3. Case (ii) Select the path with SRDP using Maximum Link Bandwidth [$M_{BW}(v_i, v_j)$ ].
   If more than one path has same Bandwidth
   Then Go to case 7 for selecting path with maximum HopsCount-Bandwidth-Delay Product (HBDP)
4. Case (iii) Select the path with SRDP using Minimum Network Delay [$M_{ND}(v_i, v_j)$ ].
   If more than one path has same Network Delay
   Then Go to case 7 for selecting path with maximum HopsCount-Bandwidth-Delay Product (HBDP)
5. Case (iv) Select the path with SRDP using Maximum HopsCount - Bandwidth Product (HBP)[$M_{HC}(v_i, v_j) * M_{BW}(v_i, v_j)$].
   If more than one path has same HopsCount - Bandwidth Product
   Then Go to case 7 for selecting path with maximum HopsCount-Bandwidth-Delay Product (HBDP)
6. Case (v) Select the path with SRDP using Maximum Bandwidth-Delay Product (BDP) [$M_{BW}(v_i, v_j) * M_{ND}(v_i, v_j)$].
   If more than one path has same Bandwidth – Delay Product
   Then Go to case 7 for selecting path with maximum HopsCount-Bandwidth-Delay Product (HBDP)
7. Case (vi) Select the path with SRDP using Minimum HopsCount- Delay Product (HDP) [$M_{HC}(v_i, v_j) * M_{ND}(v_i, v_j)$].
   If more than one path has same HopsCount- Delay Product
   Then Go to case 7 for selecting path with maximum HopsCount-Bandwidth-Delay Product (HBDP)
8. Case (vii) Select the path with SRDP using HopsCount- Bandwidth- Delay Product (HBDP) [$M_{HC}(v_i, v_j) * M_{ND}(v_i, v_j)$].
   If more than one path have same HopsCount- Bandwidth- Delay Product
   Then Select path with maximum Bandwidth-Delay Product Upper Bound (BDP UB)

## Algorithm2: Secure Optimum Route Maintenance (S -ORM)

Route Maintenance Algorithm between Broker and Cloud ($v_i$ and $v_j$)

1. Select secure communication link between broker and cloud by ORCF algorithm, which is implemented through SRDP over S-RDN.
2. Start Cloudlet transmission on the selected secure communication link.
3. Maintain list of intermediate nodes (in S-RDN) in communication link at the time of communication.
4. Check the status of link after a unit time interval.

5. If an intermediate node that detects a link break due to lack of response from a neighboring node

    Then

        i. Node sends a Route Error Packet [REP] to broker.
        ii. Broker stops forwarding cloudlets.
        iii. Broker initiate to search for another communication link by sending RRP and select link with S-ORCF.

6. Check the BDP/ UP-BDP of link after a unit time interval.
7. If BDP is decreased after unit time interval

    Then

    Broker initiates to search for another secure communication link by sending RREQ and select link with S-ORCF.

## 5. CONCLUSIONS AND FUTURE WORK

In this paper, we propose an improved BCCP with integration of security issues with S-ORCF and S-ORM algorithms using Secure Dynamic Source Routing Approach that uses symmetric cryptographic primitives. It is attained by directing every node to retain a reliable One-hop RDN and providing a secret to each node residing in the RDN and broadcast encrypted message that is not accessible by any other node that are not in the RDN. Also there are KDSs for pair wise authentication. We discussed secure route discovery protocol that employs a security paradigm in cloud computing. Performance can also be analysis based on resource utilization cost using S-ORCF and S-ORM algorithms to improve service delivery. S-ORCF & S-ORM algorithm can be analysis over cloudsim tool. It can also employ to expand an efficient Cloud DataCenter Management system, to analyze other CloudSim class files. It requires in developing new policies, algorithms, protocols for development of fast and logical network structure. We will attempt to apply Bandwidth Provisioning, Memory Provisioning; SAN Storage etc. in future work. We will analyze the performance based on resource utilization cost to reduce overall cost in service delivery through cloud.

## Authors

Gaurav Raj

M.Tech(SE) from MNNIT, Allahabad
e-mail ID- er.gaurav.raj@gmail.com
 Affiliated with Lovely Professional University, Phagwara as Assistant Professor.
Research area: Cloud Computing & Wireless Network

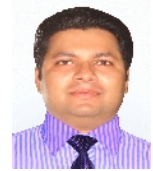

Kamaljit Kaur

M. Tech(CS) from GNDU, Amritsar (Gold Medal)
e-mail ID- kamal.aujla86@gmail.com
Affiliated with Lovely Professional University, Phagwara as Assistant Professor.
Research area: Cloud Computing & Wireless Network

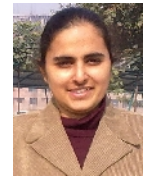